\documentclass[aps,twocolumn,prd,showpacs,showkeys,preprintnumbers,superscriptaddress,nobibnotes,floatfix,longbibliography,notitlepage,nofootinbib]{revtex4-2}

\pdfoutput=1
\usepackage{amsmath}
\usepackage{amsfonts}
\usepackage{amssymb}
\usepackage{mathrsfs}
\usepackage{color}
\usepackage{slashed}
\usepackage{graphicx}
\usepackage{xcolor}
\usepackage{hyperref}
\hypersetup{colorlinks=true,linkcolor=blue}
\usepackage{multirow, makecell}
\usepackage{listings}
\usepackage{upgreek}
\usepackage{bm}
\usepackage[capitalise]{cleveref}
\usepackage[colorinlistoftodos]{todonotes}
\usepackage[caption=false]{subfig}
\usepackage[normalem]{ulem}
\usepackage{siunitx}
\usepackage{booktabs}
\usepackage{tabularx}
	
\DeclareSIUnit \s {\second}
\DeclareSIUnit \ns {\nano\second}
\DeclareSIUnit \mus {\micro\second}
\DeclareSIUnit \ms {\milli\second}
\DeclareSIUnit \MB {\mega\byte}
\DeclareSIUnit \GB {\giga\byte}
\DeclareSIUnit \TB {\tera\byte}
\DeclareSIUnit \PB {\peta\byte}
\DeclareSIUnit \Mbps {\mega\bit/\s}
\DeclareSIUnit \Gbps {\giga\bit/\s}
\DeclareSIUnit \Tbps {\tera\bit/\s}
\DeclareSIUnit \Pbps {\peta\bit/\s}
\DeclareSIUnit \kton {\kilo\tonne} 
\DeclareSIUnit \kt {\kilo\tonne}
\DeclareSIUnit \Mt {\mega\tonne}
\DeclareSIUnit \eV {\electronvolt}
\DeclareSIUnit \keV {\kilo\electronvolt}
\DeclareSIUnit \MeV {\mega\electronvolt}
\DeclareSIUnit \GeV {\giga\electronvolt}
\DeclareSIUnit \PeV {\peta\electronvolt}
\DeclareSIUnit \EeV {\exa\electronvolt}
\DeclareSIUnit \ZeV {\zetta\electronvolt}
\DeclareSIUnit \m {\meter}
\DeclareSIUnit \cm {\centi\meter}
\DeclareSIUnit \in {\inchcommand}
\DeclareSIUnit \km {\kilo\meter}
\DeclareSIUnit \kV {\kilo\volt}
\DeclareSIUnit \kW {\kilo\watt}
\DeclareSIUnit \MW {\mega\watt}
\DeclareSIUnit \MHz {\mega\hertz}
\DeclareSIUnit \mrad {\milli\radian}
\DeclareSIUnit \year {years}
\DeclareSIUnit \POT {POT}
\DeclareSIUnit \sig {$\sigma$}
\DeclareSIUnit\parsec{pc}
\DeclareSIUnit\lightyear{ly}
\DeclareSIUnit\foot{ft}
\DeclareSIUnit\ft{ft}
\DeclareSIUnit \ppb{ppb}
\DeclareSIUnit \ppt{ppt}
\DeclareSIUnit \samples{S}
\DeclareSIUnit \pe{PE}
\DeclareSIUnit \mwe{mwe}

\newcommand{\enu}{\E_\enu}

\begin{document}

\title{MiniBooNE and MicroBooNE Combined Fit to a 3+1 Sterile Neutrino Scenario}

\author{
        A.~A. Aguilar-Arevalo$^{14}$,
        B.~C.~Brown$^{5}$, 
        J.~M.~Conrad$^{13}$,
        R.~Dharmapalan$^{1,7}$, 
        A.~Diaz$^{13}$,
        Z.~Djurcic$^{2}$, 
        D.~A.~Finley$^{5}$, 
        R.~Ford$^{5}$,
        G.~T.~Garvey$^{10}$,
        S.~Gollapinni$^{10}$,
        A.~Hourlier$^{13}$, 
        E.-C.~Huang$^{10}$,
        N.~W.~Kamp$^{13}$, 
        G.~Karagiorgi$^{4}$, 
        T.~Katori$^{12}$,
        T.~Kobilarcik$^{5}$, 
        K.~Lin$^{4,10}$,
        W.~C.~Louis$^{10}$, 
        C.~Mariani$^{16}$, 
        W.~Marsh$^{5}$,
        G.~B.~Mills$^{10,\dagger}$,
        J.~Mirabal-Martinez$^{10}$, 
        C.~D.~Moore$^{5}$, 
        R.~H.~Nelson$^{3,\star}$, 
        J.~Nowak$^{9}$,
        Z.~Pavlovic$^{5}$, 
        H.~Ray$^{6}$, 
        B.~P.~Roe$^{15}$,
        A.~D.~Russell$^{5}$,
	A.~Schneider$^{13}$,
        M.~H.~Shaevitz$^{4}$,
        J.~Spitz$^{15}$, 
        I.~Stancu$^{1}$,
	R.~Tayloe$^{8}$,
        R.~T.~Thornton$^{10}$, 
        M.~Tzanov$^{3,11}$,
        R.~G.~Van~de~Water$^{10}$,
        D.~H.~White$^{10,\dagger}$, 
        E.~D.~Zimmerman$^{3}$ \\
(The MiniBooNE Collaboration)
}
\affiliation{
$^1$University of Alabama; Tuscaloosa, AL 35487, USA \\
$^2$Argonne National Laboratory; Argonne, IL 60439, USA \\
$^3$University of Colorado; Boulder, CO 80309, USA \\
$^4$Columbia University; New York, NY 10027, USA \\
$^5$Fermi National Accelerator Laboratory; Batavia, IL 60510, USA \\
$^6$University of Florida; Gainesville, FL 32611, USA \\
$^7$University of Hawaii, Manoa; Honolulu, HI 96822, USA \\
$^8$Indiana University; Bloomington, IN 47405, USA \\
$^9$Lancaster University; Lancaster LA1 4YB, UK \\
$^{10}$Los Alamos National Laboratory; Los Alamos, NM 87545, USA \\
$^{11}$Louisiana State University; Baton Rouge, LA 70803, USA \\
$^{12}$King's College London; London WC2R 2LS, UK \\
$^{13}$Massachusetts Institute of Technology; Cambridge, MA 02139, USA \\
$^{14}$Instituto de Ciencias Nucleares; Universidad Nacional Aut\'onoma de M\'exico; CDMX 04510, M\'exico \\
$^{15}$University of Michigan; Ann Arbor, MI 48109, USA \\
$^{16}$Center for Neutrino Physics; Virginia Tech; Blacksburg, VA 24061, USA \\
$^\star$Now at The Aerospace Corporation, Los Angeles, CA 90009, USA \\
$^\dagger$Deceased \\
}

\date{\today}

\newcommand\mbdm{\SI{0.191}{\square\eV}}
\newcommand\mbdmbare{0.191}
\newcommand\mbue{0.021}
\newcommand\mbuu{0.500}
\newcommand\mbsin{0.0417}
\newcommand\mbchi{27.8}
\newcommand\mbchidof{27.8 / 3}
\newcommand\mbpval{4.09\times10^{-6}}
\newcommand\mbsigma{4.6}
\newcommand\jdm{\SI{0.209}{\square\eV}}
\newcommand\jdmbare{0.209}
\newcommand\jue{0.016}
\newcommand\juu{0.500}
\newcommand\jsin{0.0316}
\newcommand\jchi{24.7}
\newcommand\jchidof{24.7 / 3}
\newcommand\jpval{1.77\times10^{-5}}
\newcommand\jsigma{4.3}
\newcommand\jdmwc{\SI{0.033}{\square\eV}}
\newcommand\jdmbarewc{0.033}
\newcommand\juewc{0.500}
\newcommand\juuwc{0.500}
\newcommand\jsinwc{1.0}
\newcommand\jchiwc{17.3}
\newcommand\jchidofwc{17.3 / 3}
\newcommand\jpvalwc{6.04\times10^{-4}}
\newcommand\jsigmawc{3.4}

\begin{abstract}
This letter presents the results from the MiniBooNE experiment within a full ``3+1'' scenario where one sterile neutrino is introduced to the three-active-neutrino picture.
In addition to electron-neutrino appearance at short-baselines, this scenario also allows for disappearance of the muon-neutrino and electron-neutrino fluxes in the Booster Neutrino Beam, which is shared by the MicroBooNE experiment.
We present the 3+1 fit to the MiniBooNE electron-(anti)neutrino and muon-(anti)neutrino data alone, and in combination with MicroBooNE electron-neutrino data.
The best-fit parameters of the combined fit with the exclusive CCQE analysis (inclusive analysis) are $\Delta m^2 = \jdm~(\jdmwc)$, $|U_{e4}|^2 = \jue~(\juewc)$, $|U_{\mu 4}|^2 = \juu~(\juuwc)$, and $\sin^2(2\theta_{\mu e})=\jsin~(\jsinwc)$.
Comparing the no-oscillation scenario to the 3+1 model, the data prefer the 3+1 model with a $\Delta \chi^2/\text{dof} = \jchidof~(\jchidofwc)$, a $\jsigma\sigma~(\jsigmawc\sigma)$ preference assuming the asymptotic approximation given by Wilks' theorem.
\end{abstract}

\maketitle

\section*{Introduction}

The MiniBooNE low-energy excess (LEE) is a long-standing anomaly in neutrino physics.
This excess of electron-like events was observed in the muon-neutrino dominated flux from the Booster Neutrino Beam (BNB), and is most significant between $\SI{200}\MeV$ and $\SI{600}\MeV$ in reconstructed neutrino energy.
Initially reported in 2007~\cite{MiniBooNE:2007uho}, the excess reached a significance of $4.8\sigma$ in the energy range $\SI{200}\MeV < E_\nu^\text{QE} < \SI{1250}\MeV$ with the full MiniBooNE $\nu$ and $\bar \nu$ data set~\cite{MiniBooNE:2020pnu}.
We note that this significance is derived from a direct comparison between MiniBooNE data and the Standard Model (SM) prediction, and is thus independent of the any physics model, including the $3+1$ model explored in this paper.
A wide range of explanations for the excess have been put forward, but the initial, and still most-referenced, new physics explanations invoke $\nu_\mu \rightarrow \nu_e$ oscillations.

The BNB flux is produced through $\SI{8}\GeV$ protons impinging on a beryllium target that is located inside a magnetic focusing horn, which can reverse polarity to run in neutrino or antineutrino mode, followed by a $\SI{50}\m$ meson decay pipe.
The MiniBooNE detector, which is a $\SI{450}\tonne$ fiducial mass, mineral-oil-based Cherenkov detector, is located $\SI{541}\m$ downstream of the beryllium target.
The detector is sensitive to neutrinos with energies between $\SI{100}\MeV$ and $\SI{3}\GeV$.
This combination of energy and baseline makes MiniBooNE an ideal experiment to probe the appearance of electron-neutrinos from $\nu_\mu\rightarrow\nu_e$ oscillations in a mass-squared splitting region greater than $\SI{1e-2}{\square\eV}$.
The full data set taken in a series of runs between 2002 and 2019 yields a $1\sigma$ allowed region in $\Delta m^2$ between $\SI{0.04}{\square\eV}$ and $\SI{0.4}{\square\eV}$, with mixing angles varying from $1.0$ to $0.01$~\cite{MiniBooNE:2020pnu}.

These mass-squared splittings are more than an order of magnitude larger than the splitting of atmospheric neutrino oscillations, $\Delta m^2_{atmos}\approx \SI{2.5e-3}{\square\eV}$~\cite{ParticleDataGroup:2020ssz}--associated with the largest mass splitting in three neutrino oscillation models.
Therefore, to accommodate such oscillations, it is necessary to postulate the existence of a fourth neutrino mass, and a fourth neutrino flavor that must be non-weakly-interacting (or ``sterile'') to avoid constraints from $Z$ decay~\cite{ALEPH:2005ab}.
In such a model, the sterile neutrino flavor and the three active flavors are connected to a fourth mass state through an extension of the PMNS mixing matrix.
Such a model introduces a combination of three possible experimental signatures: 1) electron flavor disappearance to other flavors, leading to fewer $\nu_e$ events than expected (``$\nu_e\rightarrow\nu_e$"); 2) muon flavor disappearance to other flavors (``$\nu_\mu\rightarrow\nu_\mu$") reducing the $\nu_\mu$ rate; and 3) $\nu_\mu\rightarrow\nu_e$ appearance, where an excess of $\nu_e$ events would be observed.
Past MiniBooNE $\nu_\mu\rightarrow\nu_e$ appearance analyses have assumed that the $\nu_e$ and $\nu_\mu$ disappearance effects were negligible.
However, for the mass squared splitting and mixing angles we are concerned with, $\nu_e$ disappearance can reduce the intrinsic $\nu_e$ background contribution by up to $\SI{80}\percent$, and $\nu_\mu$ disappearance can decrease the $\nu_\mu$ event rate by up to $\SI{80}\percent$.
Neglecting these effects has been considered to be an overly simplified approach.
External analyzers have investigated the difference between this approach and an analysis with a full treatment of the 3+1 model~\cite{Brdar:2021cgb,Dentler:2018sju,Kopp:2013vaa}, and have explored the effects other nuclear models on these results~\cite{Brdar:2021cgb}.
In response to this, in this paper we expand the analyses of the full MiniBooNE data sets and simulation samples, to present the first full 3+1 sterile-neutrino oscillation model by the collaboration.

In 2015, the MicroBooNE experiment joined the MiniBooNE experiment as a user of the BNB beamline.
The MicroBooNE experiment was designed with the primary goal of investigating the LEE by using the detailed information from its liquid-argon time-projection-chamber (LArTPC) to distinguish between electron induced events and photon induced events.
This allows the rejection of many mis-identified backgrounds in the MiniBooNE data set.
MicroBooNE has recently released results of a search for a generic $\nu_e$ excess, assuming the median shape of the MiniBooNE excess, in a strategy that is agnostic to particular oscillation models.
External analyses have applied more focused studies, placing limits on $\nu_e$ disappearance~\cite{Denton:2021czb}, expanding the MicroBooNE analysis to all systematically allowed shapes of the MiniBooNE excess~\cite{Arguelles:2021meu}, and considering how the MicroBooNE data constrain the parameters of a 3+1 sterile neutrino model~\cite{Arguelles:2021meu}.
However, until now, there has been no MiniBooNE-MicroBooNE combined analysis.

Because MiniBooNE and MicroBooNE share the same beamline, we can use MiniBooNE tools to perform a combined fit to the data of the two experiments.
On the other hand, because the detectors are substantially different, the two experiments have complementary capabilities.
MicroBooNE is an $\SI{85}\tonne$ active mass LArTPC~\cite{MicroBooNE:2016pwy}, which allows for detailed reconstruction of neutrino interactions that is not possible using the MiniBooNE Cherenkov detector.

The MiniBooNE experiment has a large sample size, but relatively high backgrounds from mis-identification backgrounds that dominate MiniBooNE's electron neutrino sample.
The MicroBooNE experiment uses a relatively small detector, but can remove most mis-identification backgrounds~\cite{MicroBooNE:2021jwr}.
The imaging capability of the LArTPC has allowed the MicroBooNE experiment to select three $\nu_e$ charged-current (CC) samples~\cite{MicroBooNE:2021rmx}: a high purity exclusive sample of charged current quasi-elastic (CCQE) interactions~\cite{MicroBooNE:2021jwr}, a semi-inclusive sample of pion-less interactions~\cite{MicroBooNE:2021sne}, and a fully-inclusive sample~\cite{MicroBooNE:2021nxr}.
The MicroBooNE data allow for a clean test of the hypothesis that the MiniBooNE excess events are due to $\nu_e$ CC interactions.
We consider only the CCQE and inclusive analyses, as the former has the lowest systematic uncertainty and the latter has the largest sample size.
The CCQE analysis uses deep-learning-based reconstruction while the inclusive analysis uses Wire-Cell-based reconstruction; thus they are hereafter identified in figures and tables by the shorthand ``DL'' and ``WC'', respectively.
In this paper, we present the first MiniBooNE/MicroBooNE combined fits to a $3+1$ model.

\section*{Fit Details}
The model of interest is a three-active plus one-sterile neutrino model called ``3+1.''
This model expands the $3\times 3$ neutrino mixing matrix to $4\times 4$: 
\begin{equation}
U_{3+1} = \begin{bmatrix}
U_{e1} & U_{e2} & U_{e3} & U_{e4} \\ 
U_{\mu 1} & U_{\mu 2} & U_{\mu 3} & U_{\mu 4} \\
U_{\tau 1} & U_{\tau 2} & U_{\tau 3} & U_{\tau4} \\
U_{s1} & U_{s2} & U_{s3} & U_{s4}
\end{bmatrix}. \label{4mixmx}
\end{equation}
In such a model, both $\nu_\mu$ and $\nu_e$ disappearance are expected to occur with the same $\Delta m^2$ as the $\nu_\mu \rightarrow \nu_e$ appearance signal, as long as both $U_{e4}$ and $U_{\mu 4}$ are non-zero.
The three processes are related through their effective mixing angles, which are expressed as:
\begin{eqnarray}
\sin^2 (2\theta_{\mu \mu })&=& 4(1-|U_{\mu4}|^2)|U_{\mu4}|^2, \nonumber \\
\sin^2 (2\theta_{ee})&=& 4(1-|U_{e4}|^2)|U_{e4}|^2, \nonumber \\
\sin^2 (2\theta_{e \mu}) &=& 4 |U_{e4}|^2 |U_{\mu 4}|^2,
\label{mixings}
\end{eqnarray}
which appear within the oscillation probability formulae:
\begin{eqnarray}
P(\nu_\mu \to \nu_e) &=&\sin^2 2\theta_{\mu e}  \sin^2 (\Delta m_{41}^2 L/E), \nonumber \\
P(\nu_{e} \to \nu_e)  &=& 1-\sin^2 2\theta_{ee}  \sin^2 (\Delta m_{41}^2 L/E), \nonumber
\\
P(\nu_{\mu} \to \nu_\mu)  &=& 1-\sin^2 2\theta_{\mu \mu}  \sin^2 (\Delta m_{41}^2 L/E).
\label{osc}
\end{eqnarray}
There are three physics parameters in the 3+1 model relevant to these two experiments: the sterile mass splitting $\Delta m_{4i}^2 \equiv \Delta m^2$ (where we assume degeneracy for $i \in \{1,2,3\}$) and the two mixings of the new mass eigenstate to the electron weak eigenstate $|U_{e4}|^2$ and muon weak eigenstate $|U_{\mu4}|^2$.
Different combinations of these parameters will induce different rates of $\nu_e$ appearance as well as $\nu_\mu$ and $\nu_e$ disappearance in the MiniBooNE and MicroBooNE detectors.
In each case the oscillation probability depends upon the true neutrino energy, $E$, and baseline of each event, $L$.

\begin{figure*}
     \subfloat[\label{fig:mb_mu_ex}MiniBooNE $\nu_\mu$ + $\bar{\nu}_\mu$]{
         \includegraphics[width=0.4\linewidth]{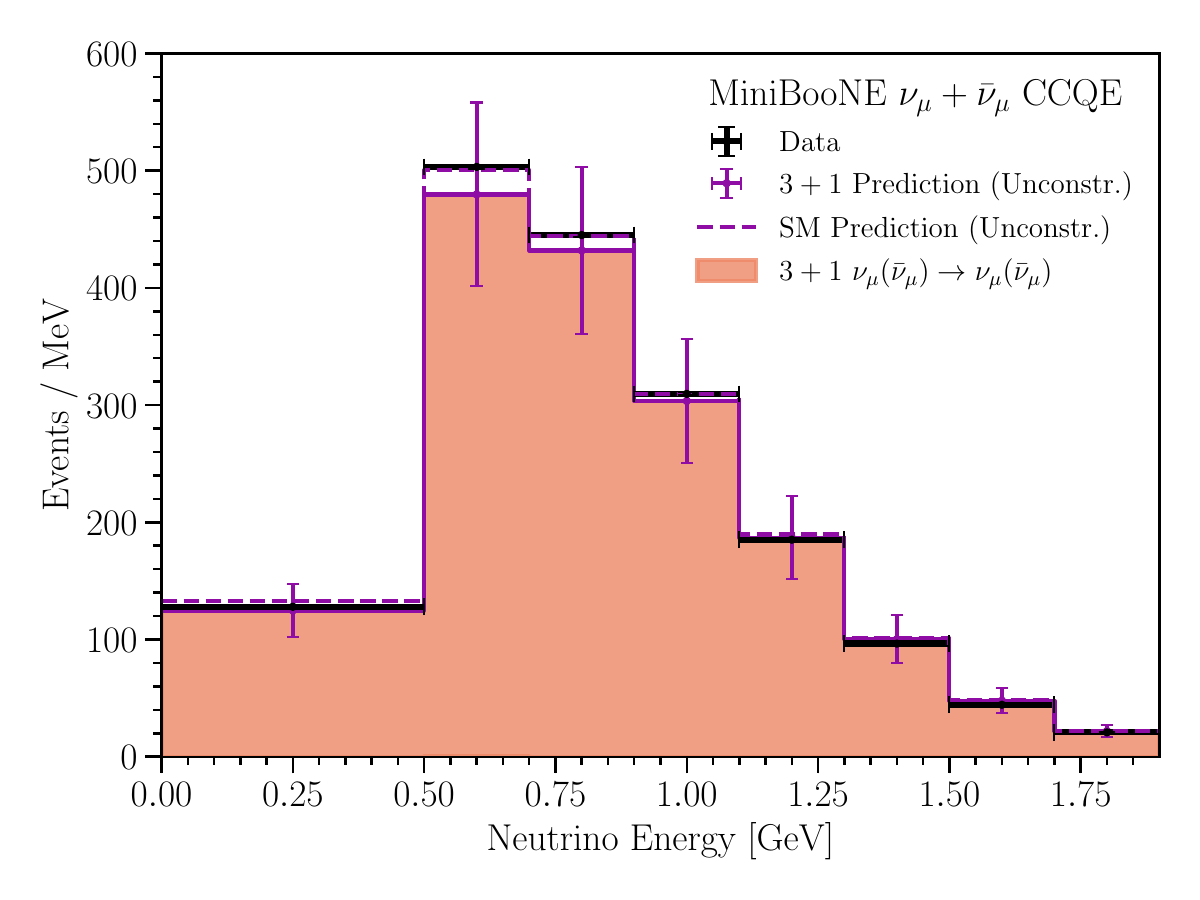}}
     \subfloat[\label{fig:mb_e_ex}MiniBooNE $\nu_e$ + $\bar{\nu}_e$]{
         \includegraphics[width=0.4\linewidth]{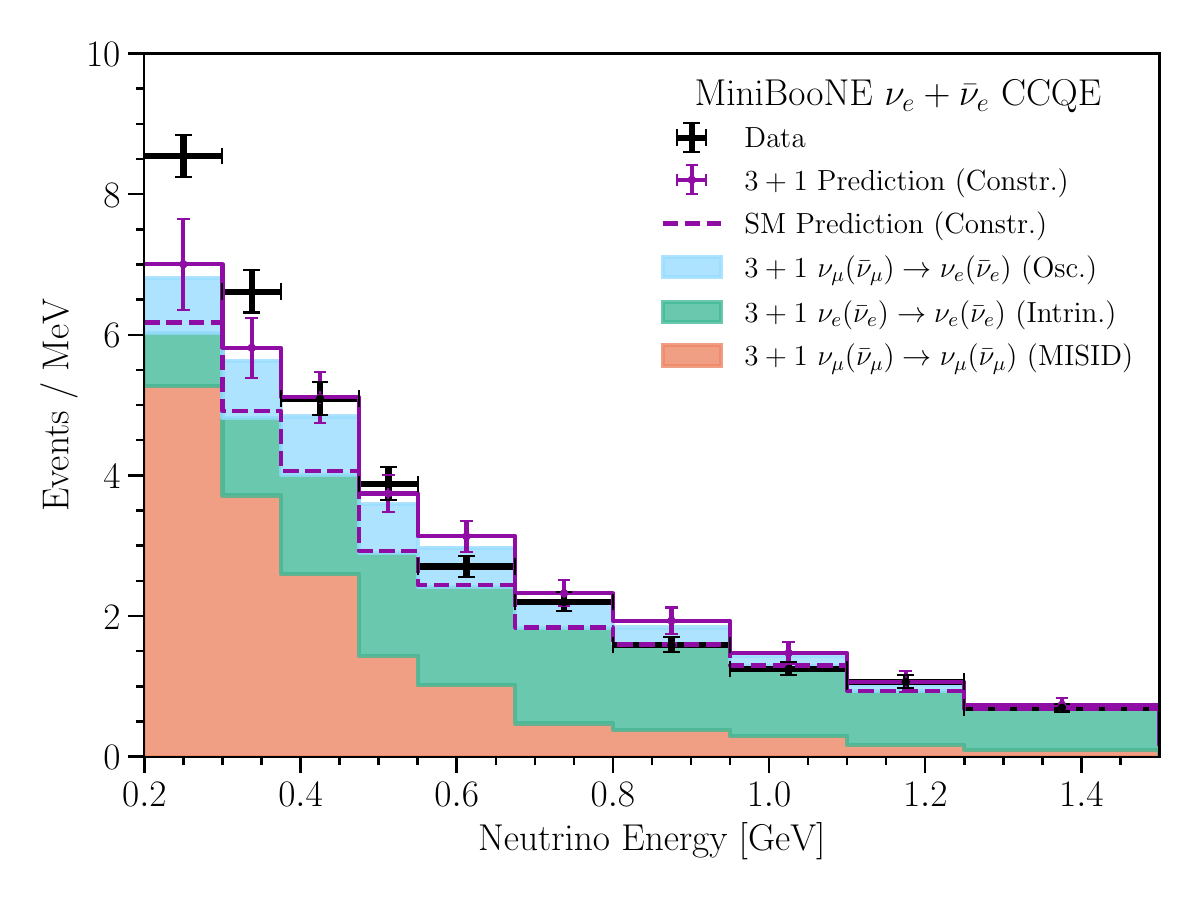}} \\
     \subfloat[\label{fig:dl_ex}MicroBooNE $\nu_e$ CCQE]{
         \includegraphics[width=0.4\linewidth]{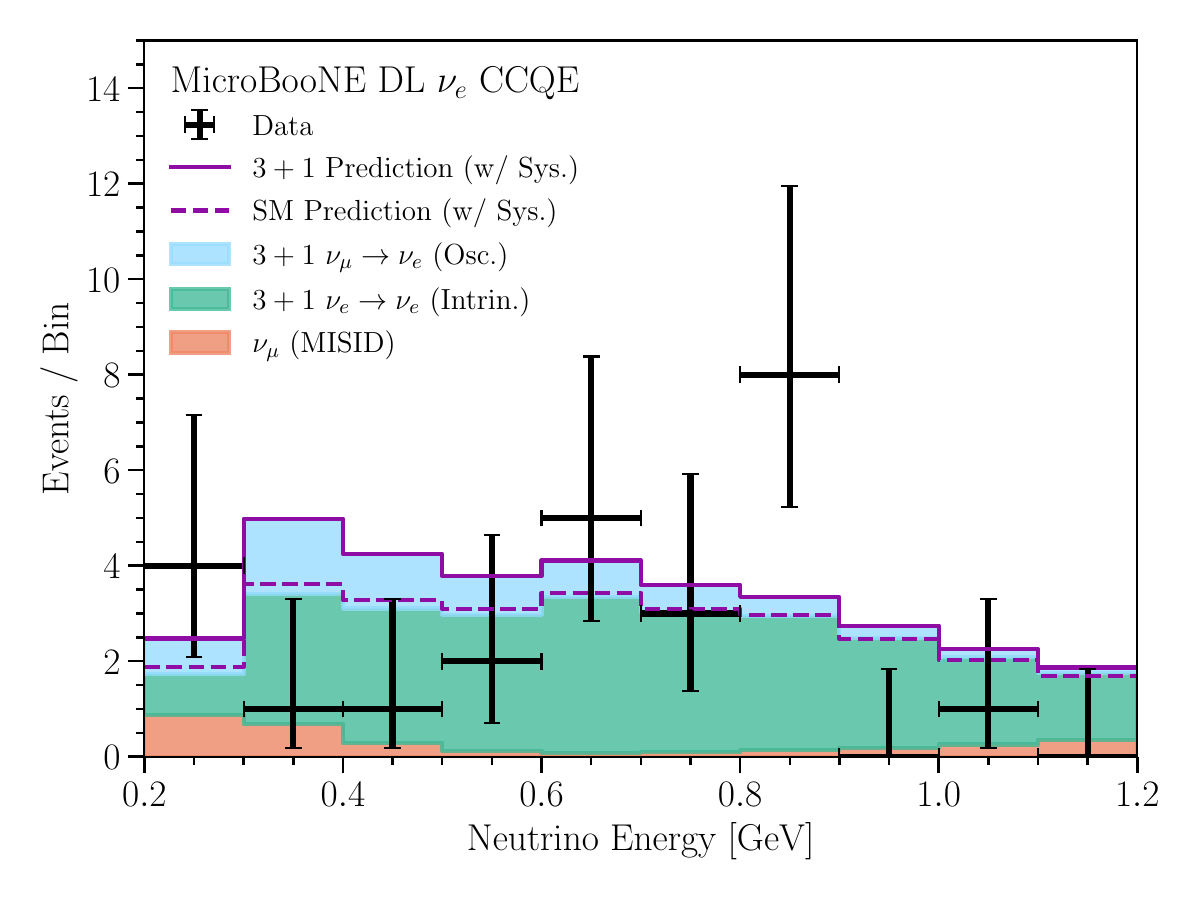}}
    \subfloat[\label{fig:wc_ex}MicroBooNE $\nu_e$ CC FC Inclusive]{
         \includegraphics[width=0.4\linewidth]{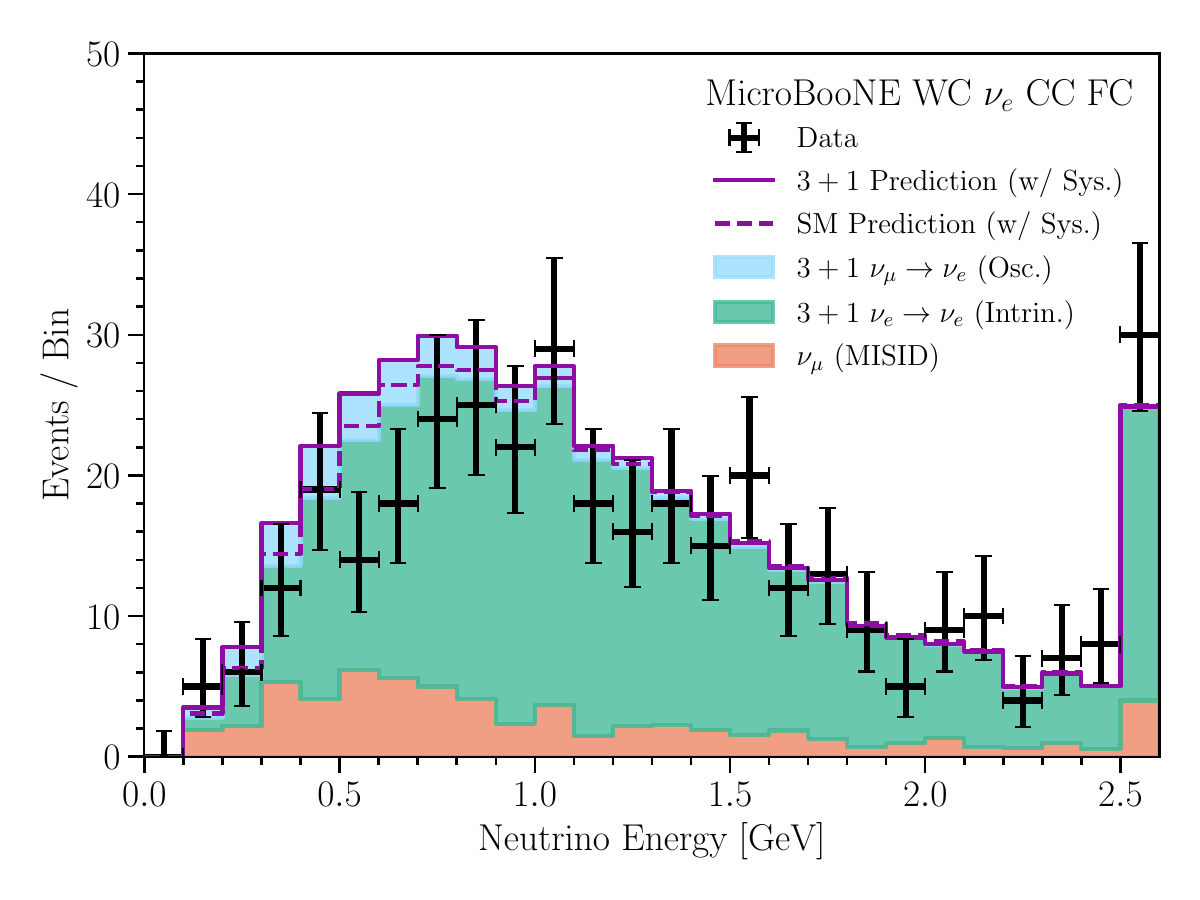}}
     \caption{Comparison between data and prediction for each experiment, showing the prediction from both the SM and the 3+1 model. The MiniBooNE and MicroBooNE DL figures (top, bottom left) consider the 3+1 ``Combination (DL)'' fit parameters of Table~\ref{tbl:bestfit}, while the MicroBooNE WC figure (bottom right) considers the ``Combination (WC)'' fit parameters.
     Black crosses show the observed data and statistical error, and stacked histograms show the unconstrained prediction.
     The SM (3+1) prediction is represented as a dashed (solid) purple line. 
     The error bars on the MiniBooNE 3+1 prediction represent systematic uncertainty.
     The top left panel shows unconstrained predictions and errors.
     The top right panel shows predictions and errors in purple after being constrained by the $\nu_\mu+\bar\nu_\mu$ data.
     The bottom panels show predictions after the allowed systematic variations have been fit to data, and thus do not have systematic error bars shown.
     }
     \label{fig:osc_ex}
\end{figure*}

The oscillation prediction in MiniBooNE is determined by a simple reweighting of the MiniBooNE $\nu_\mu\rightarrow\nu_e$ simulation using the oscillation formulae (Eqs.~\ref{osc}).
This direct method is not possible for the MicroBooNE analyses, as only limited simulation information for each analysis is available~\cite{hepdata:1953568,hepdata.114862.v2}.
Instead, for MicroBooNE, we use the MiniBooNE BNB simulation to obtain a ratio between the nominal intrinsic $\nu_e$ background prediction and the $\nu_e$ appearance prediction at the MicroBooNE baseline as a function of true neutrino energy, using the BNB flux prediction at the MicroBooNE location.
This ratio, combined with the intrinsic $\nu_e$ simulation provided by MicroBooNE allows us to obtain a $\nu_e$ appearance prediction in MicroBooNE.
We use the same procedure to account for $\nu_e$ disappearance in both analyses and $\nu_\mu$ disappearance in the inclusive analysis.
We neglect $\nu_\mu$ disappearance in the MicroBooNE $1e1p$ CCQE prediction, as the $\nu_\mu$ background contamination in MicroBooNE's $1e1p$ analysis is sub-dominant and the simulation information for the $\nu_\mu$ contribution is not provided by MicroBooNE.
In the MicroBooNE inclusive analysis, we consider only four of the seven channels: the $\nu_e$ and $\nu_\mu$ CC fully-contained (FC) and partially-contained (PC) samples.
This is because energy reconstruction information is not provided for the remaining for the remaining three $\pi^0$-based samples.
We also note that $\nu_\mu \to \nu_\tau$ neutral-current backgrounds in MiniBooNE's electron neutrino measurement are not included in the prediction; however, this effect is expected to be small.
An example of this oscillation prediction is shown in Figure~\ref{fig:osc_ex}.

For the MiniBooNE likelihood we compare the fixed observation to the theoretical expectation with a multivariate normal distribution that includes systematic uncertainties, Poisson statistical uncertainties on the expectation, and finite Monte-Carlo statistical uncertainties.
With the large MiniBooNE sample size, the multivariate normal distribution is a reasonable approximation for the likelihood.
The MiniBooNE systematic errors of this analysis remain the same as in~\cite{MiniBooNE:2020pnu}, with one exception.
The correlated systematic errors from uncertainties in the MiniBooNE optical model are limited to the three principal components of the corresponding covariance matrix with the largest eigenvalues, and the remaining optical model errors are assumed to be uncorrelated with no covariance among energy bins.
For each MicroBooNE analysis, we use a Poisson-derived likelihood that accounts for finite Monte-Carlo size~\cite{Arguelles:2019izp}; additionally, the expectation in each bin is treated as a nuisance parameter that is constrained by the systematics covariance matrix~\cite{hepdata:1953568/t3,hepdata.114862.v2/t2}.
The total likelihood is then composed of these two experimental likelihoods.
We note that although the same beamline simulation is used to derive systematic uncertainties for both experiments, because of technical limitations the fit presented here accounts for these uncertainties as if they were uncorrelated between the two experiments.
The inclusion of information from the $\nu_\mu$ samples of both MiniBooNE and MicroBooNE indirectly constrains the $\nu_e$ predictions of the two experiments in a correlated manner.
This is handled directly for the inclusive analysis, as the $\nu_\mu$ CC FC/PC samples are included in the fit (accounting for $\nu_\mu$ appearance). 
For the CCQE analysis, we allow the MicroBooNE $\nu_\mu$ $1\mu1p$ measurement to constrain the MicroBooNE $\nu_e$ $1e1p$ prediction and uncertainties, and do not account for oscillations in MicroBooNE's $\nu_\mu$ $1\mu1p$ prediction.
Ignoring $\nu_\mu$ disappearance in MicroBooNE's CCQE analysis is a reasonable assumption for small $U_{\mu4}$ given the limited sample size from this analysis.

\section*{Results}

With the methods described in the preceding section we can examine the MiniBooNE LEE in the context of a 3+1 sterile neutrino model, both with the MiniBooNE data alone and together with the MicroBooNE electron-neutrino data.
We show the no-oscillation SM prediction as a dashed purple line in Figure~\ref{fig:osc_ex}.
In the SM case the MiniBooNE prediction lies substantially below the data in the electron-neutrino channel.
For the MicroBooNE CCQE analysis, the data lie scattered above and below the SM prediction, in part due to the small sample size.
For the MicroBooNE inclusive analysis, the data lie below the SM prediction across most of the energy range.
The disparity between the data and SM prediction in MiniBooNE shows the inability of the SM to accommodate the MiniBooNE low energy excess in the electron-neutrino data while remaining in agreement with the MiniBooNE muon-neutrino data.

In contrast to the SM, the 3+1 oscillation model provides the additional freedom necessary to potentially better accommodate the MiniBooNE muon neutrino data, and low energy excess, within systematic errors.
In the 3+1 scenario we expect $\nu_\mu\rightarrow\nu_e$ oscillations to increase the prediction in the electron-neutrino channels of both experiments, while $\nu_e$ disappearance will reduce the intrinsic electron-neutrino backgrounds, and $\nu_\mu$ disappearance will reduce the muon-neutrino prediction as well as the contribution of misidentified events in the electron-neutrino observable channel.
The prediction for the best-fit 3+1 scenario across both experiments is shown in Figure~\ref{fig:osc_ex}, separated by component, experiment, and observable channel.
Figure~\ref{fig:mb_mu_ex} compares the MiniBooNE unconstrained muon neutrino and antineutrino prediction to observed data, where the crosses denote the unconstrained 3+1 prediction and the dashed line denotes the unconstrained SM prediction; here the 3+1 prediction is approximately $\SI{4}\percent$ lower than the SM prediction in the bin with the largest expectation.
Figure~\ref{fig:mb_e_ex} compares the MiniBooNE electron neutrino and antineutrino prediction to data; the prediction and errors are shown after being constrained by the muon neutrino data for the $3+1$ and SM scenarios in purple, whereas the unconstrained $3+1$ prediction is shown by the stacked histogram.
The best-fit electron neutrino 3+1 oscillation prediction is approximately $\SI{15}\percent$ lower in the lowest energy bin than that reported for the best-fit in the two neutrino oscillation analysis~\cite{MiniBooNE:2020pnu}.

While the best-fit 3+1 scenario is preferred to the no-oscillation scenario, it still cannot perfectly describe MiniBooNE's low energy excess, especially at the lowest energies.
This is consistent with the recent MicroBooNE results, which indicate that the low energy excess cannot be explained entirely by electron neutrinos~\cite{MicroBooNE:2021rmx}.
This is also consistent with previous MiniBooNE studies indicating a forward-peaked angular distribution of the low energy excess~\cite{MiniBooNE:2020pnu}.

\begin{figure}[t]
    \centering
    \includegraphics[width=\linewidth]{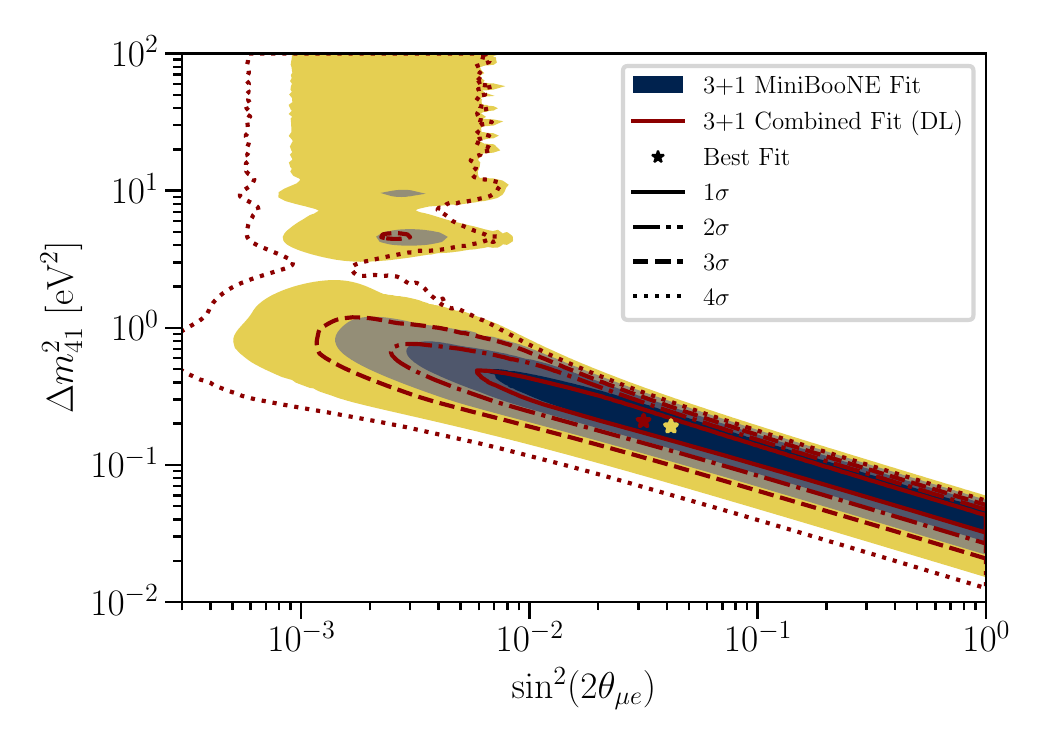}
    \includegraphics[width=\linewidth]{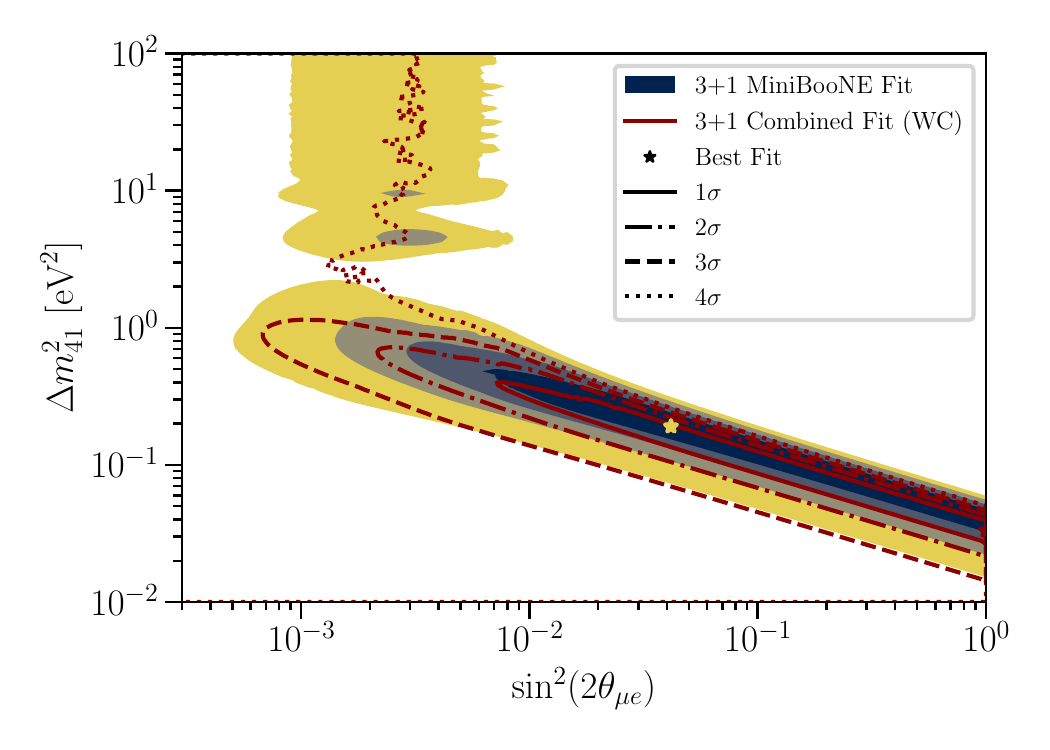}
    \caption{The results of the MiniBooNE-only and combined fits with MicroBooNE's CCQE sample~\cite{MicroBooNE:2021jwr} (top) and MicroBooNE's Inclusive sample~\cite{MicroBooNE:2021nxr} (bottom).
    The likelihood is obtained by profiling over all parameters except $\Delta m^2$ and $\sin^2(2\theta_{\mu e})$.
    The two best-fit points are shown as appropriately colored stars, and the contours are obtained by comparing the profile-likelihood-ratio test-statistic to the asymptotic distribution provided by Wilks' theorem, and assuming a difference of two degrees of freedom.}
    \label{fig:fit_contours}
\end{figure}

The best-fit 3+1 parameters and the $\Delta\chi^2$ between the SM and 3+1 scenarios are given in Table~\ref{tbl:bestfit}.
We obtain a best-fit that includes substantial sterile-muon mixing, with $|U_{\mu 4}|^2$ near $0.5$, and moderate sterile-electron mixing, with $|U_{e4}|^2$ near $0.02$, for both the MiniBooNE only and the CCQE combined fit.
For these fits the best-fit $\Delta m^2$ is near $\SI{0.2}{\square\eV}$ as well.
The large sterile-muon mixing at the best-fit point is in tension with constraints on unitarity from some experiments in the neutrino sector~\cite{Hu:2020oba,Ellis:2020hus,Parke:2015goa}, although a substantial number of neutrino experiments violate these constraints~\cite{Barinov:2021asz,Goldhagen:2021kxe,Berryman:2020agd,Kaether:2010ag,SAGE:2009eeu} as discussed in~\cite{Denton:2021czb}.
However, in this analysis, a broad region in parameter space is allowed within the estimated $1\sigma$ confidence region, as is visualized in Figure~\ref{fig:fit_contours}, extending to regions of parameter space which are not in tension with unitarity constraints.
The $1\sigma$ allowed region in $\Delta m^2$ and $\sin^2(2\theta_{\mu e})$ is similar to that reported in~\cite{MiniBooNE:2020pnu}, and takes the form of a diagonal band because the MiniBooNE LEE spans a broad energy range and extends down to the $\SI{200}\MeV$ boundary.
The inclusive search combined fit obtains a best-fit at maximal mixing, with a $\Delta m^2$ of $\jdmwc$, compatible with this diagonal band.
The excess drives the allowed values of $\sin^2(2\theta_{\mu e})$, but large deviations from the best-fit in $|U_{e4}|^2$ and $|U_{\mu 4}|^2$ are allowed, provided the combination produces enough $\nu_\mu\rightarrow\nu_e$ appearance to describe the excess.
This freedom is present in part because the systematic errors of the prediction allow large changes to the muon-neutrino channel with little penalty, which in turn provides only a weak constraint on $|U_{\mu 4}|^2$ through $\nu_\mu$ disappearance.

In both analyses, MicroBooNE's electron-neutrino data do not exhibit an excess at the lower end of their energy spectrum, as MiniBooNE's electron-neutrino data do, and MicroBooNE overall observes a lower event rate than predicted by the nominal no-oscillation model~\cite{MicroBooNE:2021rmx}.
However, the data sample from MicroBooNE does not have the statistical power needed to rule out a 3+1 $\nu_\mu\rightarrow\nu_e$ explanation of the MiniBooNE low-energy-excess.
The observed event-rate from MicroBooNE's $\nu_e$ CCQE $1e1p$ analysis precludes very large $\nu_\mu\rightarrow\nu_e$ appearance at values of $\Delta m^2$ and $\sin^2(2\theta_{\mu e})$ higher than the MiniBooNE allowed region.
This manifests in Figure~\ref{fig:fit_contours} (top) as a small shift in the allowed region to lower $\Delta m^2$ and lower $\sin^2(2\theta_{\mu e})$.
In Figure~\ref{fig:dl_ex}, the best-fit 3+1 oscillation prediction increases the expected number of events in a region where the MicroBooNE CCQE analysis observes a deficit, suggesting that the fit is primarily driven by the larger MiniBooNE data sample, in line with our expectation.
This is also true for the MicroBooNE inclusive analysis, as shown in Figure~\ref{fig:wc_ex}.
However, the inclusive analysis provides a stronger constraint on $\nu_\mu \to \nu_e$ appearance in general.
This manifests in Figure~\ref{fig:fit_contours} as (1) a more significant modification of the allowed regions for the combined fit and (2) a smaller $\Delta \chi^2$ between the 3+1 best fit and the SM in Table~\ref{tbl:bestfit}, in comparison to the combined fit with the MicroBooNE CCQE analysis.

The 3+1 scenario is preferred over the no-oscillation model in both the MiniBooNE-only and combined-fit cases.
In the MiniBooNE-only fit we obtain a $\Delta \chi^2=\mbchi$ between the two models, whereas in the combined-fit we obtain a $\Delta \chi^2=\jchi$ for 3 additional degrees of freedom introduced in the fit.
This is smaller than the $\Delta \chi^2=29$ with 3 degrees of freedom reported in the two neutrino oscillation analysis~\cite{MiniBooNE:2020pnu}, representing a drop in the significance when disappearance effects are accounted for.
If we assume the asymptotic approximation to the test-statistic distribution provided by Wilks' theorem~\cite{wilks1938} with a difference of three degrees of freedom between the models, then we obtain p-values of $\mbpval$ and $\jpval$ in favor of the 3+1 scenario for the MiniBooNE-only and combined analyses, respectively.
However, we expect the true difference in degrees of freedom between the models to be less than three, based on both the degeneracy inherent in the 3+1 model and the smaller difference in degrees of freedom observed in the two-neutrino MiniBooNE oscillation study~\cite[\S5]{MiniBooNE:2020pnu}.
A reduction in the difference in degrees of freedom between the models would increase the significance of these two statistical tests.
Therefore, we conservatively estimate that the MiniBooNE-only 3+1 model test prefers the 3+1 model to the SM at approximately $\mbsigma\sigma$, and the addition of the MicroBooNE electron-neutrino CCQE (inclusive) data reduces this significance to approximately $\jsigma\sigma$ ($\jsigmawc\sigma$).
\\

\section*{Conclusion}

\begin{table}[t!]
\newcolumntype{Y}{>{\centering\arraybackslash}X}
\begin{center}
\begin{tabularx}{\linewidth}{l Y Y Y Y}
\toprule
3+1 Fit & $|U_{e4}|^2$ &  $|U_{\mu 4}|^2$ & $\Delta m^2$ & $\Delta \chi^2$/ dof  \\
\midrule
MiniBooNE only & \mbue & \mbuu & \mbdmbare & \mbchidof \\
Combination (DL) & \jue & \juu & \jdmbare & \jchidof \\
Combination (WC) & \juewc & \juuwc & \jdmbarewc & \jchidofwc \\
\bottomrule
\end{tabularx}
\caption{Summary of results. The $\Delta \chi^2/\text{dof}$ in the last column compares the $3+1$ model to the no-oscillation model.}
\label{tbl:bestfit}
\end{center}
\end{table}

This letter has explored a full 3+1 sterile-neutrino oscillation model within the context of results from the MiniBooNE and MicroBooNE experiments.
In the MiniBooNE electron-like analysis, we consider $\nu_\mu\rightarrow\nu_e$ appearance alongside both $\nu_e$ and $\nu_\mu$ disappearance.
In the MicroBooNE CCQE analysis, we consider $\nu_e$ appearance and $\nu_e$ disappearance.
In the MicroBooNE inclusive analysis, we consider $\nu_e$ appearance and both $\nu_e$ and $\nu_\mu$ disappearance.
In an analysis of the MiniBooNE-only data, we find a best-fit to the 3+1 model of $\Delta m^2 = \mbdm$, $|U_{e4}|^2=\mbue$, $|U_{\mu 4}|^2=\mbuu$, and $\sin^2(2\theta_{\mu e})=\mbsin$.
A combined-fit to the MiniBooNE and MicroBooNE CCQE analyses finds a best fit to the 3+1 model at oscillation parameters of $\Delta m^2 = \jdm$, $|U_{e4}|^2 = \jue$, $|U_{\mu 4}|^2 = \juu$, and $\sin^2(2\theta_{\mu e})=\jsin$.
A combined-fit to the MiniBooNE and MicroBooNE inclusive analyses finds a best fit to the 3+1 model at oscillation parameters of $\Delta m^2 = \jdmwc$, $|U_{e4}|^2 = \juewc$, $|U_{\mu 4}|^2 = \juuwc$, and $\sin^2(2\theta_{\mu e})=\jsinwc$.
In the MiniBooNE only analysis, the 3+1 scenario is preferred over the no-oscillation case with a $\Delta \chi^2$/dof of $\mbchidof$, whereas in the combined analysis with MicroBooNE CCQE (inclusive) data we obtain $\Delta \chi^2/\text{dof} = \jchidof \; (\jchidofwc)$.
Although the 3+1 model is not a perfect description of the low-energy MiniBooNE electron neutrino data, we find that a 3+1 sterile neutrino oscillation scenario is a better description of the MiniBooNE data than the no-oscillation scenario and is not in tension with MiniBooNE's muon neutrino data.
We also find that the MicroBooNE electron-neutrino data do not rule out the allowed 3+1 interpretations for the MiniBooNE data, but do reduce the significance of the result and make only a small modification to the allowed regions.
We look forward to the inclusion of additional data into this combined fit from the upcoming SBND and ICARUS experiments in order to shed more light onto the 3+1 sterile neutrino hypothesis~\cite{MicroBooNE:2015bmn}.

\bibliography{main}

\end{document}


\makeatletter
\renewcommand \thesection{S\@arabic\c@section}
\renewcommand\thetable{S\@arabic\c@table}
\renewcommand \thefigure{S\@arabic\c@figure}
\makeatother


\widetext
\begin{center}
\textbf{\large Supplemental Materials: MiniBooNE and MicroBooNE Combined Fit to a 3+1 Sterile Neutrino Scenario}
\end{center}
\setcounter{equation}{0}
\setcounter{figure}{0}
\setcounter{table}{0}
\setcounter{page}{1}
\makeatletter
\renewcommand{\theequation}{S\arabic{equation}}
\renewcommand{\thefigure}{S\arabic{figure}}
\renewcommand{\bibnumfmt}[1]{[S#1]}

\section{Likelihood} \label{app:likelihood}
The physics parameters of the model are the mass squared splitting $\Delta m^2$, electron-sterile mixing $\left|U_{e4}\right|^2$, and muon-sterile mixing $\left|U_{\mu 4}\right|^2$.
The mixing parameters ($\left|U_{e4}\right|^2$, $\left|U_{\mu 4}\right|^2$) are allowed to vary between $0$ and $1$ while maintaining unitarity of the mixing matrix through the condition $\left|U_{e4}\right|^2 + \left|U_{\mu 4}\right|^2 \leq 1$.
The additional nuisance parameters of the model are the MicroBooNE per-bin systematic scalings $\alpha_i$.
Here the set of physics parameters are denoted by $\vec{\theta}$, and the set of nuisance parameters denoted by $\vec{\eta}$.
The combined MiniBooNE-MicroBooNE likelihood is the product the two experimental likelihoods such that
\[
\mathcal{L}(\vec{\theta},\vec{\eta}|\vec{x}) = \mathcal{L}_\text{MB}(\vec{\theta}|\vec{x}_\text{MB}) \times \mathcal{L}_\text{uB}(\vec{\theta},\vec{\eta}|\vec{x}_\text{uB}),
\]
where $\mathcal{L}_\text{MB}$ is the MiniBooNE likelihood, $\mathcal{L}_\text{uB}$ is the MicroBooNE likelihood, $\vec{x}_\text{MB}$ is collection of the MiniBooNE data counts, $\vec{x}_\text{uB}$ is the collection of MicroBooNE data counts, and $\vec{x}=\vec{x}_\text{MB}\cup\vec{x}_\text{uB}$ is the collection of all data counts.
The MiniBooNE likelihood is approximated as a multivariate normal distribution
\[
\mathcal{L}_\text{MB}(\vec{\theta},\vec{\eta}|\vec{x}_\text{MB}) = \mathcal{N}(\vec{x}_\text{MB}|\vec{\mu}_\text{MB}(\vec{\theta}),\bm{\Sigma}_\text{MB}(\vec{\theta})),
\]
where $\vec{\mu}_\text{MB}$ is the predicted number of data counts in each bin, and $\bm{\Sigma}_\text{MB}$ is the MiniBooNE covariance matrix.
In this case the MiniBooNE covariance matrix includes systematic errors, Poisson statistical errors, and Monte-Carlo statistical errors.
The MicroBooNE likelihood is given by
\[
\mathcal{L}_\text{uB}(\vec{\theta},\vec{\eta}|\vec{x}_\text{uB}) = \mathcal{N}(\vec{\alpha}|1,\bm{\Sigma}_\text{uB}) \times \prod_i \mathcal{L}^\text{Eff}(\alpha_i\mu_i^\text{uB}(\vec{\theta}),\sigma_{i,\text{mc}}^2(\vec{\theta},\alpha_i)|x_{i,\text{uB}}),
\]
where $\mathcal{N}(\vec{\alpha}|1,\bm{\Sigma}_\text{uB})$ is the multivariate normal prior on the MicroBooNE systematics scalings, $\bm{\Sigma}_\text{uB}$ is the MicroBooNE fractional covariance matrix, $\mu_i$ is the predicted number of data counts in each bin before systematic modifications, and $\sigma_{i,\text{mc}}^2$ is the Monte-Carlo statistical error on the per-bin data count prediction after the systematics scalings have been applied.
The MicroBooNE fractional covariance matrix, $\bm{\Sigma}_\text{uB}$, is the constrained fractional covariance matrix from~\cite{MicroBooNE:2021jwr,hepdata:1953568/t3}.
The likelihood $\mathcal{L}^\text{Eff}$ is a Poisson-based likelihood that accounts for finite Monte-Carlo sample errors, and is described in~\cite{Arguelles:2019izp}. The minimum $-\log\mathcal{L}$ for each of the 3+1 fit scenarios described in the main text is given in Table~\ref{tab:likelihoods}.

\begin{table}[h]
    \centering
    \begin{tabular}{| c | c | c |}
    \hline
    Datasets Included & $\min_{\vec{\eta}}\{ -\log\mathcal{L}\}$ (no-oscillation model) & $\min_{\vec{\theta},\vec{\eta}} \{ -\log\mathcal{L}\}$ (3+1 model) \\
    \hline
    MiniBooNE only & 219.1 & 205.2 \\
    MiniBooNE + DL & 251.0 & 238.7 \\
    MiniBooNE + WC & 673.1 & 664.5 \\
    \hline
    \end{tabular}
    \caption{Summary of the minimum negative-log-likelihood values for each of the 3+1 fit scenarios described in the main text, including the result for both the no-oscillation and oscillation cases.}
    \label{tab:likelihoods}
\end{table}

\section{Combined Fit with MicroBooNE Inclusive Analysis} \label{app:inclusive}

\begin{figure*}
     \subfloat[MiniBooNE $\nu_\mu$ + $\bar{\nu}_\mu$]{
         \includegraphics[width=0.4\linewidth]{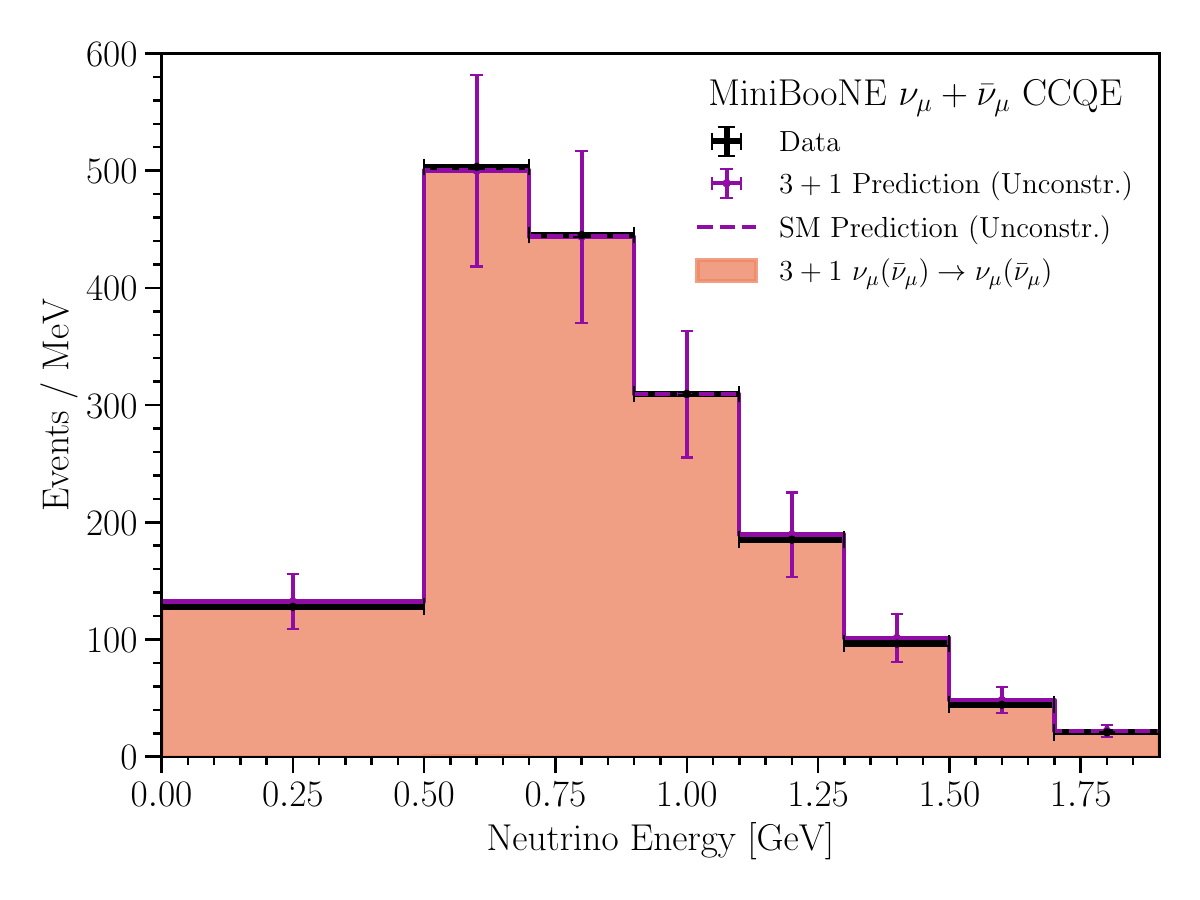}}
     \subfloat[MiniBooNE $\nu_e$ + $\bar{\nu}_e$]{
         \includegraphics[width=0.4\linewidth]{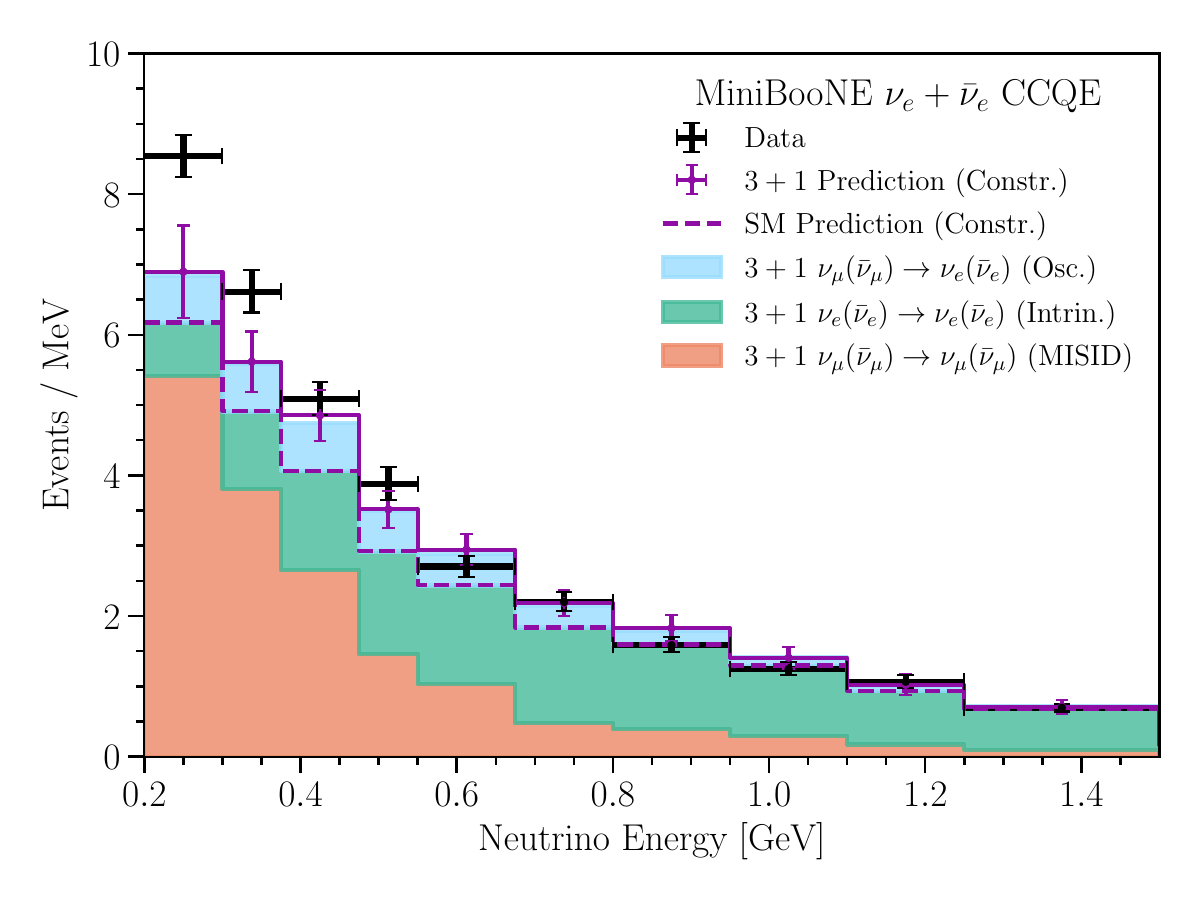}} \\
     \subfloat[MicroBooNE $\nu_e$ CC FC Inclusive]{
         \includegraphics[width=0.4\linewidth]{JointWC_MicroBooNE_WC_nueFC.pdf}}
    \subfloat[MicroBooNE $\nu_e$ CC PC Inclusive]{
         \includegraphics[width=0.4\linewidth]{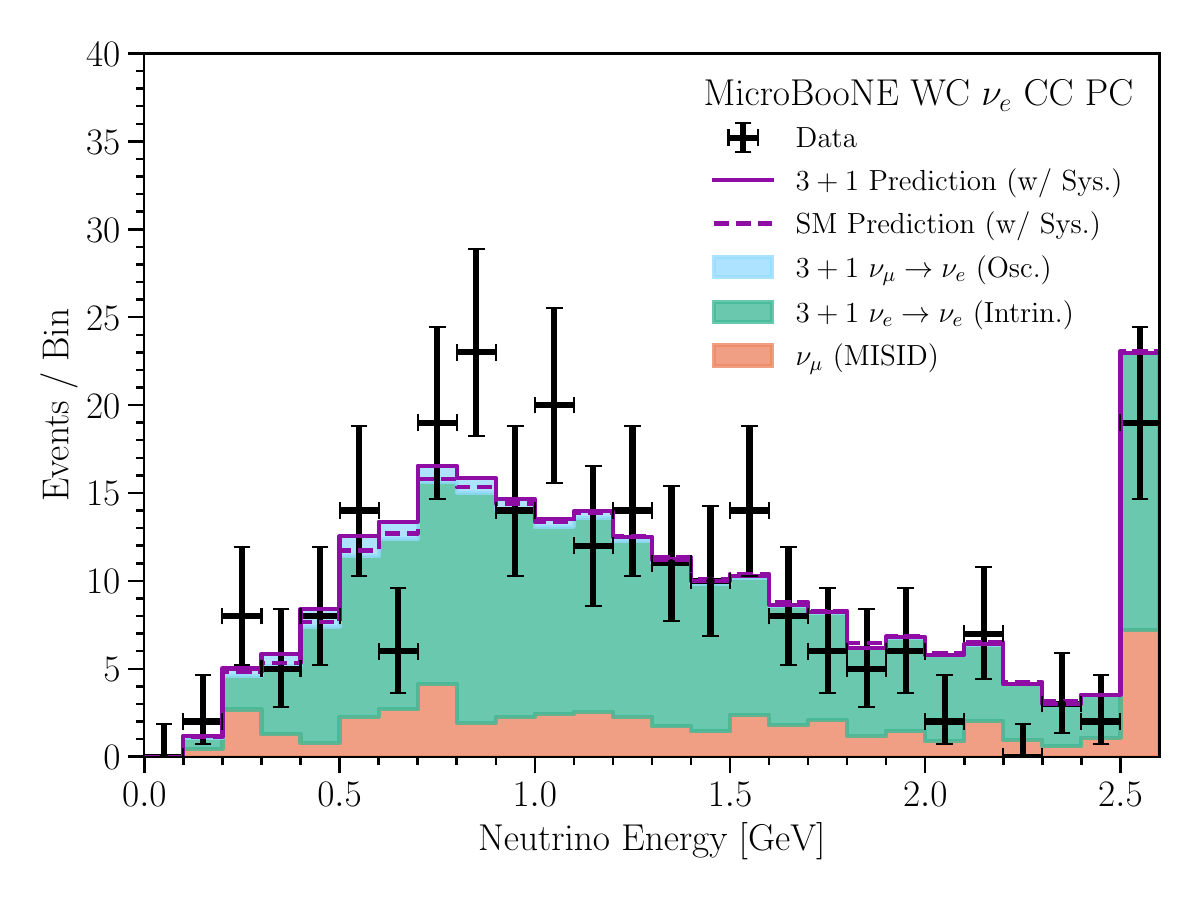}}
         \\
    \subfloat[MicroBooNE $\nu_\mu$ CC FC Inclusive]{
         \includegraphics[width=0.4\linewidth]{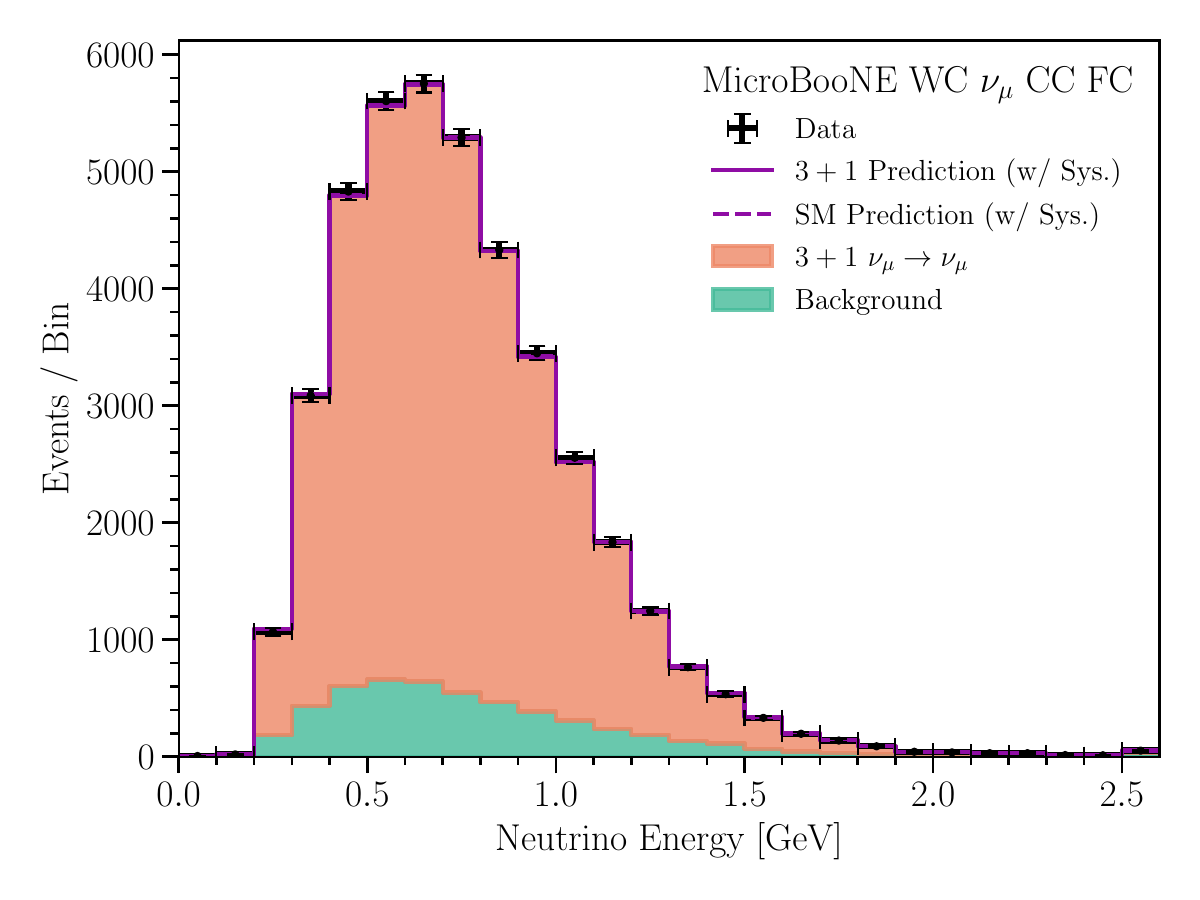}}
    \subfloat[MicroBooNE $\nu_\mu$ CC PC Inclusive]{
         \includegraphics[width=0.4\linewidth]{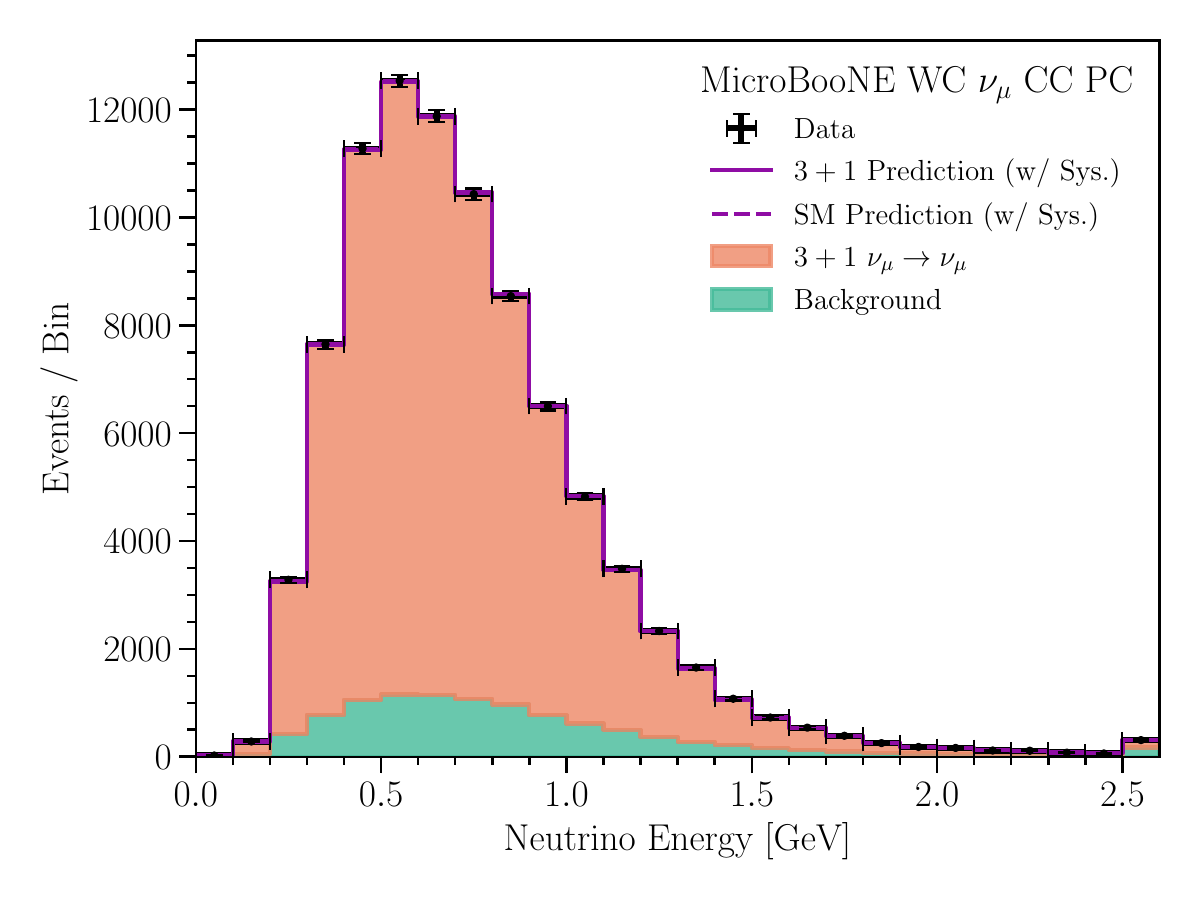}}
         \\
     \caption{Comparison between data and prediction for each experiment, showing the prediction from both the SM and the 3+1 model, considering the 3+1 ``Combination (WC)'' fit parameters of Table~I. See the caption of Figure~1 in the main text for a description of the content of each subfigure.
     }
     \label{fig:wc_osc_ex}
\end{figure*}

In this section, we provide the predicted event event rate in the MiniBooNE $\nu_e + \bar{\nu}_e$ and $\nu_\mu + \bar{\nu}_\mu$ distributions, as well as well as the MicroBooNE $\nu_e$~FC, $\nu_e$~PC, $\nu_\mu$~FC and $\nu_\mu$~PC distributions for the ``Combination (WC)'' best fit to the 3+1 model in Table~I.
Comparisons between data and prediction in each of these channels is shown in Figure~\ref{fig:wc_osc_ex}.
One can see that the best-fit solution prefers negligible $\nu_\mu$ disappearance while still allowing for enough $\nu_\mu (\bar{\nu}_\mu) \to \nu_e (\bar{\nu}_e)$ appearance to explain most of the excess in the MiniBooNE $\nu_e  + \bar{\nu}_e$ channel.
Additionally, the systematic pull terms in the MicroBooNE analysis modify the prediction in the $\nu_\mu$~FC and $\nu_\mu$~PC channels to match the data, in contrast to the central value prediction shown in Figure~21 of Ref.~\cite{MicroBooNE:2021nxr}.

\section{Constraints from MicroBooNE data} \label{app:ub_only}

\begin{figure*}
     \subfloat[MicroBooNE CCQE Sample]{
         \includegraphics[width=0.5\linewidth]{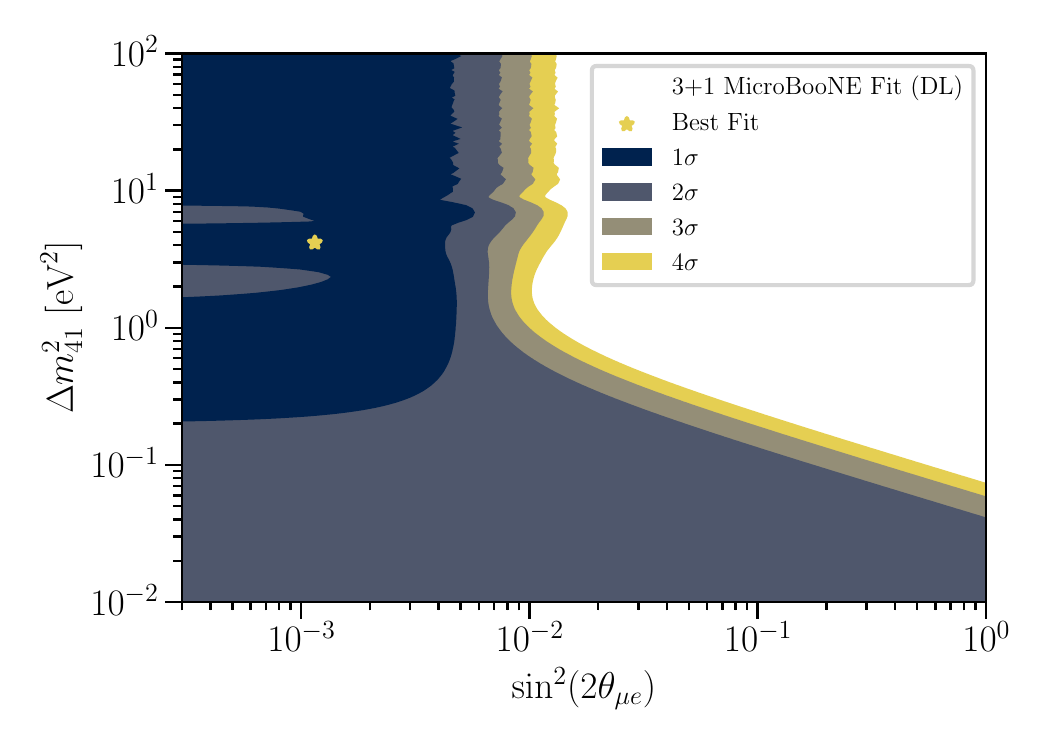}}
     \subfloat[MicroBooNE Inclusive Sample]{
         \includegraphics[width=0.5\linewidth]{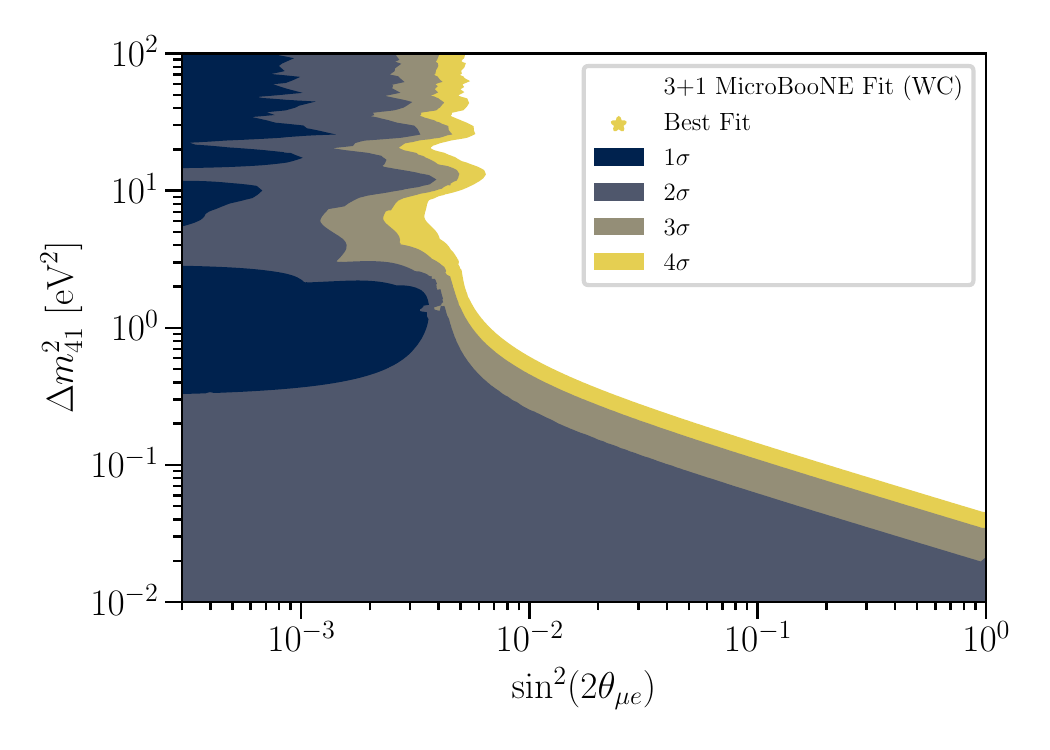}} \\
     \caption{Constraints in 3+1 parameter space from the MicroBooNE CCQE (left) and Inclusive (right) samples. See the caption of Figure~2 in the main text for more details on the content shown in each plot. Note that the best fit point for the Inclusive sample lies outside of the window shown here.
     }
     \label{fig:ub_only_osc}
\end{figure*}

In this section, we report the constraints in 3+1 parameter space derived from each of MicroBooNE samples individually.
Results from the CCQE (Inclusive) sample are shown in the left (right) plot of Figure~\ref{fig:ub_only_osc}.
One can see that the Inclusive sample sets a slightly stronger constraint than the CCQE sample.

\bibliography{main}